\documentstyle[12pt,aasms4]{article}

\lefthead{C. Gibson and R. Schild}
\righthead{Clumps of hydrogenous planetoids as
the dark matter of galaxies}

\begin{document}
 
\title{Clumps of hydrogenous planetoids as the
dark matter of galaxies}
 
\author{Carl H. Gibson}
\affil{Center for Astrophysics and Space Sciences\\ University
of California,
   San Diego, CA 92093-0411 \\ cgibson@ucsd.edu}

\and
 
\author{Rudolph E. Schild}
\affil{Center for Astrophysics,
    60 Garden Street, Cambridge, MA 02138;
\\ rschild@cfa.harvard.edu}

\begin{abstract}  Hydrodynamic gravitational
condensation theory and quasar-microlensing
observations lead to the conclusion that the
baryonic mass of most galaxies is dominated by
dense clumps of hydrogenous planetoids.  Star
microlensing collaborations fail to detect
planetoids as the dominant dark matter component
of the inner Galaxy halo (within $\approx$ 30
kpc) by an unjustified  uniform-number-density
assumption that underestimates the average value.
At plasma neutralization and photon decoupling,
existing proto-galaxies should fragment at both
proto-globular-cluster (PGC) $\approx 10^5
M_{\sun}$  and  terrestrial-mass
scales $\approx 10^{-6}
M_{\sun}$, from Gibson's 1996 hydro-gravitational
theory.  Schild's 1996 interpretation was that
the mass of the lens galaxy is dominated by
``rogue planets ... likely to be the missing
mass'',  from measured twinkling frequencies of
the lensed quasar Q0957+561 A,B images and their
time-delayed difference. 
Schild's findings of a 1.1 year image time delay
with dominant planetoid  image-twinkling-period
are confirmed herein by three observatories.  
\end{abstract}


\section{Introduction}  We discuss an
accumulation of observational evidence and 
theoretical predictions supporting the conclusions
that the masses of galaxies within about
$10^{21}$ m (30 kpc) of their central cores are
dominated by Proto-Globular-star-Cluster (PGC)
mass ($10^{35}$ kg $\--$ $10^{5}M_{\sun}$) 
clumps of planetary mass ($10^{24}$ kg $\--$
$10^{-6}M_{\sun}$)  objects formed from
primordial hydrogen-helium gas soon after its
transition from plasma $300 \, 000$ years after
the Big Bang.  The planetoids are termed
``rogue planets'' by \cite{sch96} and
``primordial fog particles'' (PFPs) by
\cite{gib96}. 

The most convincing observational evidence comes
from light curves of gravitationally lensed
quasars.  A few quasars have galaxies precisely
along their lines of sight.  The overall mass
distributions of such galaxies serve as lenses
that distort space and cause the quasar to appear
as two or more brightened, twinkling, images. The
dominant mass component of the lens galaxy
determines the dominant twinkle period (or
equivalently, the peak frequency of the microlensing light
curve spectrum) of the quasar-image light curves,
from Einstein's gravitational equations, as
various objects pass in front of the quasar at
their estimated transverse velocity.  The smaller the
object mass, the shorter the period of
brightening or darkening (the twinkle period).
\cite{kee82} was the first to use such evidence
to conclude that the masses of galaxies are not
dominated by their stars or by other objects with
stellar mass.
\cite{ref93} noted,  from fluctuations of the
four images of the ``Einstein cross'' quasar
Q2237, that the mass of the lensing quasar was
possibly dominated by  objects as small as
$10^{23}$ kg ($10^{-7} M_{\sun}$),
although without image delay corrections for
intrinsic quasar variability this conclusion was
considered tentative.   

From a fifteen year record of observations of
both A and B images of the first lensed quasar
detected (Q0957+561 A,B),
\cite{sch96} and
\cite{sch97} determined the dominant twinkling
frequency of the difference between the A and B
brightness curves, corrected for an estimated
$404 \pm 26$ day time delay of image B (now
$416.3 \pm 1.7$, \cite{pel98}) to eliminate
brightness variations intrinsic to the quasar. 
\cite{sch96} concluded from the
time-delay corrected brightness difference curve,
the microlensing record, 
that the mass of the lensing galaxy of Q0957 is
dominated by component objects with planetary
mass.  A precise quasar lens time delay is required to
distinguish between intrinsic brightness changes
of the quasar and microlensing events caused by the
galaxy mass components, since typical event times
were found to be only 10 to 100 days.  The time
delay was controversial for a number of years,
but is now confirmed by several observers (\S
\ref{sect5}) supporting the \cite{sch96}
conclusion that the lens galaxy mass is dominated
by ``rogue planets ... likely to be the missing
mass''.  Although star-microlensing
collaborations fail to detect planetoids as the
halo missing mass and claim to exclude them (\S
\ref{sect2}), their search focused on larger mass 
dark matter candidate objects (brown dwarfs).

In the following section we present records from
three observatories (\S
\ref{sect5}) confirming the existence of rapid quasar microlensing
events, and confirming that the events tend to
occur in clumps.  The missing mass of the galaxy
therefore consists of primordial planetoidal objects that
have not yet accreted to form stars.  Clumping of
such objects is to be expected as a consequence
of the accretional process, with the intermittency
of planetoid number density increasing with the
mass range.  However, tight clumping or any
clumping within the clumps of planetoids poses a
measurement problem for their detection by the
star-microlensing collaborations (\S
\ref{sect2}), which have not yet detected any
planetoidal component in the halo mass of the
Milky Way Galaxy, contrary to the
quasar-microlensing observations of planetoids in
other galaxies.

Independently and simultaneously with the
\cite{sch96} observations and conclusions,
\cite{gib96} predicted that the mass of all
galaxies should be dominated by
globular-cluster-mass clumps of planetoids  based
on a new (non-Jeans) gravitational condensation
theory, discussed in the  (\S
\ref{sect1}). Details and refinements of the
new theory are given by \cite{gibs00a},
\cite{gib99}, and \cite{gib00a}b.  We summarize
our conclusions in
\S
\ref{sect6}.

\section{A quasar-microlensing event recorded at
three observatories}
\label{sect5}

A heightened interest in the Q0957 gravitational
lens system resulted from a prediction by
\cite{kun95} that a rapid decline in the quasar's
brightness should be seen in Feb.---Mar. 1996
in the second arriving B image, based upon
observation of the event in the first arriving A
image in December 1994. Because the time delay
was still controversial, with values of 1.1-years
(\cite{sch90},
\cite{pel94}) and 1.4-years (\cite{leh92},
\cite{pre92}), it appeared that observations
during February and May 1996 would settle the
time delay issue. Thus at least 3 observatories
undertook monitoring programs to observe the
predicted event.

At Mount Hopkins, the 15 year monitoring program
on the 1.2 m telescope continued, with
observations made by scheduled observers on 109
nights. Four observations were made each night
with a Kron-Cousins R filter, and the
observations averaged together for a published
nightly mean brightness value. The quasar
brightness was referenced to 5 nearby stars whose
brightnesses were checked relative to each other
to ensure stability of the magnitude zero point.
The data are plotted in Figures 1 and 2 of this
report, and data for the first season showing the
brightness drop in the first-arriving A component
have been published by \cite{sch97}.

The Princeton data were obtained by
\cite{kun95} using the 3.5m Apache Point
telescope with g and r filters on the Gunn
photometric system. Their data are not published,
but data for the first season were posted on a
World-Wide-Web site listed in the Kundic report.
We have converted these data to a standard R
filter using the relations given in \cite{ken85}.
When we compare the Princeton results to Mt.
Hopkins data for the same dates, we find an rms
residual of 0.029 mag for component A and 0.18
mag for component B. Unexpectedly, we find the
origin of this disagreement not to be so much in
random errors as in a systematic drift in the
apparent zero points in the course of the
observing season. Data for the second observing
season have not been presented in tabular form,
but a plot of the data posted at the Princeton
WWW site has allowed us to compare the results.
We have taken the Princeton data plot, separated
the two colors of data, and rescaled data for the
Mt. Hopkins and Canary Island (\cite{osc96})
groups to make the comparisons in Figures 1 and 2.

The Canary Island data are posted at the WWW site
given in the report by \cite{osc96}. They were
obtained with the 0.8m telescope using standard R
filters and local comparison stars. Because data
were obtained in response to the Princeton
challenge of \cite{kun95}, the Canary Island
group reports data for the second season only. We
have compared their data with the Mt. Hopkins
data with the assumption that any Canary Island
datum taken within 24 hours  of a Mt. Hopkins
observation had agreeing dates, and the rms
deviations of the two data sets for our 15
agreeing dates is 0.013 mag for image A and 0.016
mag for B. The error estimates listed at the WWW
web site are considerably larger, averaging 0.022
for A and 0.020 for B. Because the Canary
Island---Mt. Hopkins comparison must have some
error contribution from Mt. Hopkins, it is clear
that the posted Canary Island error estimates are
too large by a factor of approximately 2. In our
plots of the Canary Island data, Figure 2, we have
used the original posted error estimates.

We show in Figure 1 a comparison of the available
photometries for the first observing season. In
the upper plot, the Mt. Hopkins data are shown
with a magnitude scale and zero point for a
standard R filter. The Princeton data have been
shown with an arbitrary offset of 0.2 mag. These
magnitudes are determined from a transformation
from the Princeton Gunn g,r photometric system,
using the transformation equations determined by
\cite{ken85}. Error bars are shown strictly
according to the estimates of the authors. The
data are superimposed in the bottom panel of
Figure 1, and the error bars are suppressed for
clarity. It may be seen that the data agree about
as well as predicted from the errors. One
artifact that may be noticed is that there appear
to be several points, mostly in the Princeton
data, markedly below the mean trend. It is
surprising that these discrepant points are in
the sense of brightness deficiency, because the
two principal error sources, cosmic rays and
merging of the two quasar images due to bad
seeing effects, both tend to make the images
brighter. The A component data in the lower panel
will be compared to the observations of image B
in Figure 3.

In Figure 2 we show 3 data sets in the upper
panel, with their associated error bars. However
as noted previously, we do not actually have the
tabulated data for the Princeton team, and we
have scaled the results of Mt. Hopkins and the
Canary Island groups to the Princeton data as
posted in a plot released by the Princeton team.
Thus the plotted magnitudes are on the Gunn
photometric system, which differs from standard R
by a zero point offset and a color term of 0.15
mag. In other words, R = r - 0.15(g-r) + Const. 
Since image B varied only from 1.071 to 1.142, a
variation of 0.07 mag, we conclude that the
scatter introduced into the comparison of r and R
magnitudes has a full amplitude of 0.01
magnitudes, or a scatter of at most 0.005
magnitudes around a mean offset. Thus we have
simply combined the Princeton r magnitudes with
an arbitrary zero point offset in the comparison
with the Mt. Hopkins and Canary Island R
magnitudes in the bottom panel of Figure 2.

We find in Figure 2 (bottom) good evidence that
the brightness drop predicted by
\cite{kun95} did indeed occur at around Julian
Date 2450130. The brightness in the R band did
indeed drop almost 0.1 magnitudes, and time
delays of 423 days (\cite{osc97}) and 416 days
(\cite{kun97}) are determined. However, a
remarkable thing happened at the end of this
event, or immediately afterward; a strong
microlensing event was observed, principally in
the Mt. Hopkins data. The event may be seen as a
strong downward spike centered on J.D. 2450151.
Although the event was primarily seen in the Mt.
Hopkins data, the brightness did not recover to
the expected level for another 30 days, and for
the remainder of this discussion, we refer to this
event as the 3-observatory microlens.

A much better perspective on the 3-observatory
microlens comes from inspection of Figure 3,
where we plot data for both observing seasons
combined with the 416-day time delay of
\cite{kun97}. In this plot the open and filled
symbols refer to the first and second observing
seasons, exactly as in Figures 1 and 2. We
consider that from J.D. 2450151 to 2450180 the
data records are sufficiently discrepant to
conclude that a microlensing event of 30 to 40
days duration and asymmetrical profile occurred.
It is of course possible that more than one event
was occurring at this time. It is likely that
another event was seen at 2450220 $\pm$ 10 days,
again seen by 3 observatories. A few other
significant discrepancies may be recognized in
this fascinating combined data record, and it is
not surprising that the time delay has been so
difficult to determine because of the influence
of this complex pattern of microlensing. On the
other hand, with the time delay now measured, the
microlensing provides a powerful probe of the mass
distribution of objects in the lens galaxy, and
perhaps elsewhere (Schild 1996).  The mass of the
object causing the 35 day duration microlensing
event at J.D. 2450151 is $1 \times 10^{-6}
M_{\sun} = 2 \times 10^{24}$ kg.

We now pose the question of the significance
level of the detection of microlensing. We avoid
questions  of
\it{a posteriori} \rm statistics by phrasing a
test as follows. A dramatic microlensing event
was seen covering dates J.D. 2450150-70. During
the previous year, a brightness record was
obtained that covered the same time interval. If
we average and smooth the brightness record for
the previous year, at what level of statistical
significance can we say each observatory noted a
departure in the second year? Posed this way, we
can easily determine that each of the three
observatories observed a departure attributed to
microlensing of at least 10 $\sigma$, where the
standard deviation
 $\sigma$ has been estimated for the individual
data points of each observatory. Thus we conclude
that each of three observatories has obtained as
at least a 10
$\sigma$ result that the second arriving image
has brightness departures attributed to
microlensing, because they were not seen in the
first arriving image.

\section{Comparison with star microlensing
searches}
\label{sect2} The search for MAssive Compact Halo
Object (MACHO)  particles in the Halo of our 
Galaxy by microlensing stars of the Large
Magellanic Cloud (LMC)
 has resulted in negative but controversial
results.  The classic Alcock et al. (1995abcde)
papers of the MACHO  collaboration report lens
masses $m \le 0.1 M_{\sun}$,  for which event
times $t_{sm}
\approx 130
\sqrt{m/M_{\sun}}$ (days) are several months.
Star-microlensing  events by objects with PFP
mass $10^{-6} M_{\sun}$ last only 0.13 days (3
hours), and are therefore difficult to detect
in the original program which only obtained a 
single image frame of each
field in a night. The
corresponding Q0957 quasar-microlensing time
$t_{qm} \approx 3 \times 10^4
\sqrt{m/M_{\sun}}$ (days) is 30 days.

\cite{alc96} report that for a limited subsample
of their data, where several exposures of rapid
succession were considered, the low detection
rates indicate non-detection of sufficient mass
to make PFP's the entire mass of a standard
spheroidal dark matter Halo.  \cite{ren98} reach
the same conclusion from a more intensive search
of a smaller area.  The combined  MACHO and EROS
(Exp\'{e}rience de Recherche d'Objets Sombres)
collaborations (\cite{alc98}) focus on
small-planetary-mass objects such as PFPs in
excluding a population with mass 
$M_p = (10^{-7} - 10^{-3}) M_{\sun} $ as more
than 25\% of the missing halo mass within 50 kpc
of the Galaxy center (the distance to the LMC), or
$M_p = (3.7 \times 10^{-7} - 4.5
\times10^{-5}) M_{\sun} $ having more than 10\%.  
Here it is important to recognize that the star 
microlensing projects have not detected the baryonic
dark matter, but have only rejected their own model
of the dark matter and its distribution. Their assumption
that the objects are homogeneously distributed
is most unlikely for a
small-mass population such as PFPs which are
hydrogenous and primordial, and consequently
distributed as a complex array of nested clumps
due to their nonlinear
gravitational-accretion-cascade  for a wide range
of mass to form stars.  For small
$M_p$ values, the number density $n_p$ is likely
to become a lognormal random variable with
intermittency factor $I_p \equiv
\sigma^2_{ln[n_p]} 
\approx 0.5 ln [M_{\sun}/M_p] = 8.1$
(\cite{gibs00b}), where $\sigma^2_{X}$ denotes the
variance of random variable $X$.  For a
lognormal random variable, the mean to mode ratio
is $exp[3 I_p / 2] = 1.8
\times 10^{5}$ for $I_p = 8.1$.   A small number
of independent samples of $n_p$ gives an estimate
of the mode  (the most probable value) of a
random variable, which is what is estimated by
MACHO/EROS $n_p$ measurements since the LMC
occupies only about 0.04\% of the sky and the rapid
sampling required for a small object search comprised
only about 0.2\% of their records.  An exclusion of
$10^{-7} M_{\sun}$ objects as $ \le 0.1 M_{halo}$
from an estimate of the mode of
$n_p$ is thus not conclusive, since the mean PFP
halo mass  could be $ \ge 1.8 \times 10^{4}
M_{halo}$ from such measurements even if the closely
packed star samples are considered independent.
Consequently, we suggest that the
\cite{alc98} and  
\cite{ren98}  interpretations of  MACHO/EROS
statistics as an exclusion of planetoids comprising
the Halo mass are highly model dependent and as yet
inconclusive.

\section{Theory}
\label{sect1} Many astrophysical and
cosmological models of structure formation
(\cite{pad93}, \cite{pbl93}, \cite{kol94},
\cite{slk89},
\cite{win72},
\cite{ree76}) are based on the
gravitational instability criterion of
\cite{jns02}.  By this criterion, for a
homogeneous gas of density
$\rho$ and sound speed
$V_S$, the smallest possible scale of
gravitational condensation is
$ L_J \equiv V_S/(G\rho)^{1/2}$, where
$G$ is Newton's gravitational constant.  The
validity of Jeans's theory has
been questioned by Gibson (1996, 1997ab, 1998,
1999, 2000ab).  Why should the speed
of sound
$V_S$ be relevant to gravitational instability? 
Why are viscous forces and the inertial-vortex
forces of turbulent flows neglected?  What about
magnetic forces and molecular diffusivity?  The new
theory shows that fluid mechanically determined
Schwarz length scale criteria $L_{SX}$ apply rather
than
$L_J$,
\cite{gib96}. 
Condensation and void formation
are possible at lengths matching the
largest of the Schwarz scales  $L_{SX}$, where
$L_{SX}$ are derived by balancing gravitational
forces with viscous, inertial-vortex, or magnetic
forces of the fluid, or (for the super-diffusive
non-baryonic dark matter) by balancing the
diffusion velocity and gravitational velocity of
the density field,
\cite{gib99}, 2000a.  The subscript $X$ denotes
$V, T, M$ or $D$, respectively.  The Jeans theory and
its corollary misconceptions ``pressure support''
and ``thermal support'' are most misleading when
applied to the hot, quiet, early universe, when
$L_J
\gg L_{SV}
\approx  L_{SV}$ for the baryonic matter.

According to the Gibson 1996-2000 hydro-gravitational
theory, the Jeans scale
$L_J$ was irrelevant at the time of first gravitational
structure formation $t \approx 10^{12}$ s in the plasma
epoch when
$V_S = c/3^{1/2}$ was large and
$L_{SV} \approx L_{ST}
\ll ct \ll  L_J$, giving proto-supercluster to
proto-galaxy mass objects ($10^{16} M_{\sun}$). 
Recent cosmic microwave background observations of
$\delta T / T$ show a sub-horizon spectral peak
consistent with these first structures but
inconsistent with sonic interpretations of current
cosmologies.  
$L_J$  sets the formation size and mass of
proto-globular-starclusters (PGCs) for
$t
\approx 10^{13}$ s at the beginning of the gas
epoch, but
$ L_J$ should not be used as a criterion to exclude
the simultaneous formation of smaller gassy
hydrogenous planetoids (PFPs) at the viscous and
weak turbulence Schwarz scales, where 
$L_{SV} \approx L_{ST} \ll L_{J}$.  Our claim is
that an important error of current cosmological
models is to assume that the Jeans 1902 criterion
may be used under all or any circumstances to
exclude the gravitational formation of objects or
voids.  Although it is generally recognized that
either powerful turbulence or strong magnetic forces
can prevent star formation at
$L_J$ scales (\cite{cha51}); for example, in
dense molecular clouds produced by supernovas
where
$L_{ST}$ and $L_{SM}
\gg L_J$, cases of the early universe
where
$L_{SX}
\ll L_J$ have been misinterpreted or overlooked.

Jeans neglected viscous and inertial-vortex
forces in the conservation of momentum
equation with gravity. He assumed the pressure
depends only on the density.  Either of these
unjustified assumptions reduce the fluid mechanical
problem to one of gravitational acoustics.  Jeans
also neglected particle and gravitational diffusive
terms in the conservation of mass equation,
providing another important source of error.  As
shown in
\cite{gib96}, the Jeans assumptions are
inadequate to describe the highly nonlinear
process of gravitational structure formation.  The
linearized Euler equation Jeans assumed is rarely
reliable in any fluid mechanical context,
particularly to describe rapidly expanding flows
such as that of the early universe where viscous
or buoyancy forces are required to suppress
turbulence.   

Strong turbulence was not present at the time of
plasma to neutral gas transition, from
measurements of the cosmic microwave background
(CMB) radiation that show
$\delta T/T$ values of only $10^{-5}$, three or more orders
of magnitude less than values expected if the
flow were strongly turbulent.  It follows that either
the Reynolds number or Froude number or both must
have been subcritical in the plasma epoch before
$300 \, 000$ years to suppress turbulence to the
small levels indicated by the CMB, \cite{gib00a}b. 
\cite{gib99} estimates the photon viscosity of
the plasma at the time of first structure
formation was $5 \, 10^{26}$ $\rm m^2 \, s^{-1}$,
giving an horizon Reynolds number slightly above
critical and a viscous Schwarz scale mass $\rho
L_{SV}^3$ of $10^{46}$ kg, the observed mass of a
galaxy supercluster, where $L_{SV}
\equiv (\nu
\gamma / \rho G)^{1/2}$ is the viscous Schwarz
scale, $\nu$ is the kinematic viscosity,
$\gamma \approx 1/t$ is the rate-of-strain of the
fluid, and
$\rho$ is the density.  The time $t$ when the
horizon mass $\rho(t) L_{H}^3$ increases to
$10^{46}$ kg is about $10^{12}$ s, where
$\rho(t)$ is derived from Einstein's equation,
\cite{gib97b}.

As the universe
expanded and cooled the average density decreased
and the viscous condensation scale increased
slowly, giving a decrease in the condensation mass
to galactic values of about
$10^{42}$ kg by the time of plasma-gas transition
at $10^{13}$ s ($300 \, 000$ years).  The
formation of proto-supercluster and proto-galaxy
structures during the plasma epoch is a source of
buoyancy forces and subcritical Froude numbers
that can partly explain the CMB
indications of suppressed turbulence.  Because
the gravitational free fall time $ \tau_G
\equiv (\rho G)^{-1/2}$ at each fragmentation
stage in the plasma epoch exceeds the universe
age $t$, no large increase in density could
occur in these structures, although voids could
form between them.  The baryonic density of about
$10^{-17}$ kg $\rm   m^{-3}$ estimated at $t
\approx 10^{12}$ s is close to the average
density of globular star clusters, and is taken
to be a fossil of this time of first
fragmentation in the universe.

Because of their relatively small mass and
rapidity of formation (a few thousand years) from the
primordial gas, the  $L_{SV}$ scale gas
planetoids formed within PGCs are termed
``primordial fog particles,'' or PFPs.  PFP
formation represents the first gravitational
condensation of the universe, when  mass
density $\rho$ first increased due to gravity
after $300 \, 000$ years of decreasing. 
Previous structure formations in the plasma
epoch were fragmentations by void formation
because gravitational free fall times
$\tau_G \equiv (\rho G)^{-1/2}$ required for
densities to significantly
increase were longer than the age of the universe
at neutralization.  Proto-superclusters and
proto-galaxies densities have
monotonically decreased ever since, from an
initial value of $10^{-15}$ (including the
nonbaryonic component) at $t \approx 10^{12}$ s, to
present values of about
$10^{-23}$ and
$10^{-21}$ kg $\rm m^{-3}$, respectively, as the
``flat'' universe has expanded to its present average
density
$\rho \approx 10^{-26}$ kg $\rm m^{-3}$.

A thermal-acoustic-gravitational instability occurred
at the time of neutral gas formation, causing
fragmentation of the proto-galaxy gas
blobs at the Jeans scale
$L_J \approx 10^4 L_{SV}$, simultaneous to the
fragmentation at $L_{SV}$ scales to form PFPs. 
From the ideal gas equation
$p/\rho = RT \approx {V_S}^2$ it follows that
changes in the density $\rho$ are exactly
compensated by changes in the pressure $p$ if the
temperature $T$ is constant, where $R$ is the gas
constant of the hydrogen-helium mixture. Thus,
radiative heat transfer cannot inhibit void
formation at scales
$L_{IC}
\equiv (RT/\rho G)^{1/2} = L_J$, where
$L_{IC}$ is termed the initial condensation
scale (\cite{gibs00a}).  For the primordial gas
temperature
$T = 3 \, 000$ K and primordial gas density
$\rho \approx 10^{-17}$ kg $\rm m^{-3}$, the
initial fragmentation mass was $\rho L_{IC}^{3}
\approx 10^{35}$ kg, the mass of a typical
globular cluster of stars, with internal
fragmentation into a trillion PFPs.  

Observations that globular clusters typically
contain a million small, ancient, stars have
often been cited as evidence of the validity of
Jeans's theory, but we see that this
thermal-acoustic-gravitational instability has
nothing to do with the linear
perturbation stability analysis of
\cite{jns02}.  Instead, the $L_{IC}$ scale sets
the mass of increasingly isolated, $10^5 M_{\sun}$
clouds or clumps of PFPs in the expanding
proto-galaxy to that of a proto-globular-cluster, or
PGC. Relative motions of the
$10^{12}$ PFPs within the $10^{5}$ PGCs of a
proto-galaxy would have been strongly inhibited
by drag forces of their gassy environment during
their $\tau _G$ condensation periods lasting
millions of years.  Their accretion to form
globular cluster stars must have been equally
gentle, as evidenced by a remarkable spherical
symmetry which Jeans likened to ``fuzzy cricket
balls'', compared to the chaotic conditions and
geometries of present strongly turbulent star
forming regions. Old globular star clusters
provide fossil evidence of the high
density and weak turbulence existing at their
time of formation.  Young globular star clusters
with the same mass and density, \cite{ash98},
support the present hypothesis that all luminous
globular star clusters form, when tidal forces
trigger accretion, from the abundance of dark PGC
clumps of hydrogenous planetoids permeating all
galaxies.

Where are these hundreds of thousands of PGC
clumps of PFPs formed in each proto-galaxy at
the plasma-gas transition?  A full review of the
evidence is beyond the scope of the present
paper.  To summarize the model, most PGCs are
apparently intact but dark, and reside where they
were formed, in galaxy inner-halos and cores.  Some
PGCs have collided with each other or have been
otherwise agitated to activate a full formation of
stars; for example, in the 200 observed globular
clusters of the Milky Way or the $10\,000$ in M87. 
Some have been disrupted and their PFPs
re-evaporated or dispersed, possibly by repeated
encounters to form the galaxy disk and core and the
dominant mass component of interstellar medium in
these luminous spiral galaxy regions. Some may have
formed the so-called super-star clusters detected
near galaxy cores, with
$10^{4} M_{\sun}$  mass, many large
($\approx 100 M_{\sun}$) stars, and huge average
densities ($10^{-15}$ kg
$\rm m^{-3}$) requiring a compaction process. 
What provides the material
of construction for the 700 young globular
star clusters detected in the bright arms of
interacting ``antennae'' galaxies  NGC 4038/4039,
\cite{whi95}? Their typical PGC mass, density, and
size suggest the bright young PGs form from dark PGCs
brought out of cold storage by tidal forces of
the galaxy encounter.  Random gas clouds would have
random smaller densities and masses.  Why have PGC
densities remained constant?  Because the
temperature of the universe fell below the 13 K
freezing point of their gasses only after about a
billion years, the evolution of PFPs in clusters has
been gassy, complex, and collisional.  Even after
freezing to their present mode as rogue Jovian
planets they are much more subject to re-evaporation
to gas than stars, with consequent drag forces that
damp ``virial''  velocities, and hence are less
likely to be expelled from their PGC clusters by
collisionless processes than stars from star
clusters as discussed by
\cite{bin87}.  

Thus, although many questions remain, we
have good reason to expect that most of the
baryonic matter of a galaxy  consists of dark
clumps of hydrogenous planetoids, as observed by
\cite{sch96} and predicted by
\cite{gib96}.

\section{Conclusions}
\label{sect6} Quasar-microlensing evidence that
lens galaxy masses may be dominated by point-mass
objects of planetary mass has been reviewed, and
it is found that this possibility is not excluded
by the present lack of star-microlensing
evidence.  The assumption of a 
uniform-number-density of the hypothetical
population of planetoids as the halo dark matter
of the galaxy is unjustified and unexpected since
such a population should be hydrogenous and
primordial as the material of construction of
galactic stars that are observed and since any
accretion process over a million-fold mass range
is likely to produce a highly non-uniform number
density distribution for the planetoids within a
proto-globular-cluster, even if PGCs were found
on the plane of a star-microlensing field.

Because quasar images are guided by the mass of
the lens galaxy, twinkling at the dominant
point-mass-frequency is assured.  Clumping of clumps,
and subsequent clumping of point masses within
the clumps, should cause interference and multiple
point-mass-lensing with
occasional isolated events, as observed, and
render the planetoids functionally invisible to
sparse star-microlensing samples,
\cite{gibs00b}.

The \cite{sch96} inference of planetoids as the
missing matter of the Q0957+561 A,B lensed quasar
and his estimated 1.1 year time delay of the
images are confirmed by three observatories, and
supported by the
\cite{gib96} nonlinear-hydrodynamic-gravitational
condensation theory that independently predicted
this result.  Furthermore, the
\cite{gib96} theory predicts the planetoids
should be sequestered in proto-globular-clusters,
which can  explain the lack of
star-microlensing evidence of their existence,
along with the probability of strong internal
intermittency and clumping of the planetoids
within the PGCs due to their accretional cascade.

\begin{acknowledgments} 
\end{acknowledgments}

Figure 1. Data for the A (northern) gravitational
lens image, recorded in the October 1994 - June
1995 observing season. In the upper panel,
brightness estimates with error bars are shown as
triangles for the Schild and Thomson (1997) data,
and as circles for the Kundic et al. (1995) data.
The R magnitude scale is for the Schild and
Thomson data and the Kundic et al. data is
arbitrarily offset 0.2 mag to permit comparison.
In the lower panel, the data are shown
superimposed and without error bars, to show the
generally good agreement, especially around the
date of the large quasar brightness drop at
2449715. In Figures 1, 2, and 3 the most
significant 2 digits of the Julian date have been
suppressed for clarity.

Figure 2. Data for the B (southern) component
recorded in the Nov.  1995$\--$June 1996
observing season. In the upper panel, filled
squares with error bars are from Oscoz et al.
1996, Triangles are from Schild and Thomson
(1997), and circles are from the
WWW plot at the site reported by Kundic et al.
(1995).  In the lower panel, data from the three
observatories are plotted without error bars.
Generally good agreement is shown in the
comparison, and distinct brightness trends are
seen in all three data sets.

Figure 3. Data from the lower panels of Figures 1
and 2 are shown superimposed with the same symbol
definitions as previously, and for a 416 day time
delay. It may immediately be seen that there is
generally good agreement, and that the second
arriving B image (solid symbols) generally follows
the pattern of fluctuation exhibited the year
before in the first-arriving A image (open
symbols). However there are important
differences; around Julian Date 2450150-70 a
strong brightness drop occurred that had not been
seen in the first arriving A image. Similarly,
around J.D. 2450220 the records differ
systematically by several percent.

\end{document}